# Superconducting flux qubit with ferromagnetic Josephson π-junction operating at zero magnetic field

Sunmi Kim[1,*,†], Leonid V. Abdurakhimov[2,5,‡,†], Duong Pham[3,6], Wei Qiu[4,5], Hirotaka Terai[4], Sahel Ashhab[1], Shiro Saito[2], Taro Yamashita[3,6,§], and Kouichi Semba[1,7]

[1]Advanced ICT Research Institute, National Institute of Information and Communications Technology (NICT), 4-2-1, Nukui-Kitamachi, Koganei, Tokyo 184-8795, Japan
[2]NTT Basic Research Laboratories, NTT Corporation, 3-1 Morinosato-Wakamiya, Atsugi, Kanagawa 243-0198, Japan
[3]Graduate School of Engineering, Nagoya University, Furo-cho, Chikusa-ku, Nagoya, Aichi 464-8603, Japan
[4]Advanced ICT Research Institute, NICT, 588-2 Iwaoka, Nishi-ku, Kobe, Hyogo 651-2492, Japan
[5]present address: IQM Finland Oy, Keilaranta 19 02150, Espoo, Finland
[6]present address: Graduate School of Engineering, Tohoku University, Aoba 6-6-05, Aoba-ku, Sendai, 980-8579, Japan
[7]present address: Institute for Photon Science and Technology, The University of Tokyo, Tokyo 113-0033, Japan

*kimsunmi@nict.go.jp, ‡leonid.abdurakhimov@meetiqm.com,
§taro.yamashita.c8@tohoku.ac.jp
†These authors contributed equally.

## Abstract

Conventional superconducting flux qubits require the application of a precisely tuned magnetic field to set the operation point at half a flux quantum through the qubit loop, which complicates the on-chip integration of this type of device. It has been proposed that by inducing a π-phase shift in the superconducting order parameter using a precisely controlled nanoscale-thickness superconductor/ferromagnet/superconductor Josephson junction, commonly referred to as π-junction, it is possible to realize a flux qubit operating at zero magnetic flux. Here, we report the realization of a zero-flux-biased flux qubit based on three NbN/AlN/NbN Josephson junctions and a NbN/PdNi/NbN ferromagnetic π-junction. The qubit lifetime is in the microsecond range, which we argue is limited by quasiparticle excitations in the metallic ferromagnet layer. Our results pave the way for developing quantum coherent devices, including qubits and sensors, that utilize the interplay between ferromagnetism and superconductivity.

## Introduction

The essential component in superconducting quantum bits (qubits) is the Josephson junction (JJ) composed of a nanoscale tunnel barrier sandwiched between two superconducting layers. These junctions, typically formed by superconductor/insulator/superconductor structures, introduce circuit nonlinearity while preserving its quantum nature, enabling the circuit to behave as a macroscopic artificial atom. The conventional choice for these JJs, ever since the first demonstration of nanosecond-scale quantum coherent oscillations in a charge qubit in 1999[1], is the aluminum (Al)/aluminum oxide ($AlO_x$)/Al JJ. This choice is preferred due to its simplicity of fabrication using the shadow evaporation technique and its ability to provide a reliable sample quality for achieving long coherence times. Despite significant progress in improving the coherence times of Al-based qubits through advanced qubit designs[2-4], there remain challenges in terms of material improvements to deal with two-level fluctuators originating from uncontrollable defects in the amorphous $AlO_x$ in Al-based JJs[5]. Consequently, there is a growing need for materials-oriented research and design innovations to enhance the performance of superconducting qubits and reduce noise or the sensitivity of qubits to noise.

Alternative approaches have been explored to enhance device coherence. For example, titanium nitride (TiN) was used for capacitors and/or microwave resonators to mitigate microwave dielectric loss caused by uncontrolled defects in oxides present at their surfaces and interfaces[6,7]. Recently, tantalum was used for a similar purpose[8,9], approaching millisecond qubit lifetimes.

Motivated by the objective of improving materials, we have been working on the development of alternative materials not only for the capacitor and resonator components but also for the JJs in the qubits. Recently we successfully demonstrated epitaxially grown nitride superconducting qubits on a silicon (Si) substrate, achieving coherence times at the scale of tens of microseconds[10]. For the qubit design, we use a capacitively shunted (C-shunt) flux qubit[11] because of its good coherence and relatively high anharmonicity, as well as improved device-to-device reproducibility[12,13]. However, large-scale integration of flux qubits is challenging because of the requirement of an external half-flux-quantum bias to achieve the optimal flux-insensitive operation point. Considering the realistic variations in qubit loop areas, it is practically impossible to bias a large number of qubits on the same chip all at half-flux bias. If a π-junction[14-16] is incorporated in the flux qubit loops, optimal operation can be achieved at zero external flux bias, which would solve the magnetic flux uniformity problem[17-20]. Furthermore, using a qubit design whose optimal operation point is at zero flux bias is expected to help reduce magnetic noise, since any finite applied field will necessarily have fluctuations that act as noise on the qubits. This type of qubit is sometimes called quiet qubit, because it is expected to be efficiently decoupled from environmental noise, as investigated in Refs. 21 and 22.

A previous study[19] on the incorporation of ferromagnetic π-junctions in a variety of superconducting circuits illustrated the potential of these junctions in classical and quantum logic applications. Inserting a π-junction in a direct current superconducting quantum interference device (DC SQUID) demonstrated the π-phase shifting functionality of the junction. When a π-junction was embedded in a loop containing a conventional JJ that realizes a phase qubit, any extra decoherence arising from the presence of the π-junction in the loop was barely noticeable. Another study demonstrated the fabrication of a flux qubit that incorporates a π-phase shifter, achieved through hybrid Al-based JJs and a niobium (Nb)/copper-nickel (CuNi)/Nb π-junction[20]. In their pioneering work, the authors observed a shift in the magnetic field dependence of the dispersive response of a resonator coupled to the qubit, which was attributed to the π-phase shifter. However, coherent qubit operation was not demonstrated.

In this article, we present the successful operation of a flux qubit that contains a π-junction (to which we refer as π-junction qubit). The π-junction was realized using a ferromagnetic palladium-nickel ($Pd_{89}Ni_{11}$) layer between niobium nitride (NbN) superconductors[23,24]. This qubit operates optimally at zero magnetic field and demonstrates coherence times in the microsecond range, an orders-of-magnitude improvement over the nanosecond coherence times in phase qubits with π-junction[19].

## Results and discussions

**NbN-based flux qubit with/without π-junction.**

Conventional flux qubits and π-junction qubits were fabricated using NbN-based fabrication techniques, employing a TiN buffer layer on a Si substrate, as depicted in Fig. 1. Both types of qubits had the same basic design of capacitively shunted flux qubits with NbN/aluminum nitride (AlN)/NbN JJs. Two of the three JJs that comprise the flux qubit were nearly identical, and the third JJ had a smaller area by a factor $\alpha$ (< 0.5). The only difference between the qubits was the inclusion of an additional and relatively large π-phase shifter, made of a NbN/PdNi/NbN junction. Further information on the fabrication process and experiment setup can be found in "Methods".

**Microwave spectroscopy of cavity and qubits.**

In this experiment, we chose an architecture with a C-shunt flux qubit coupled to a 3-dimensional (3D) microwave cavity[13]. The reason for using the 3D cavity is that it offers a well-controlled electromagnetic environment for the qubit, avoiding the unintended interaction with spurious microwave modes found in 2-dimensional resonators[25,26], and there is also the benefit of reduced surface dielectric losses through 3D qubit designs[27]. For example, $T_1$ times up to 90 μs were observed in Al-based C-shunt flux qubits on a sapphire substrate in a 3D cavity[13]. Details on the cavity design and characterization can be found elsewhere[13].

The qubits were placed inside the cavity as shown in Fig. 1a and characterized at a base temperature of 10 mK in a dilution refrigerator. The qubits were coupled to the lowest-frequency TE101 mode of the cavity[13]. To confirm the coupling between the qubit and the cavity, we measured the response of the microwave transmission ($S_{21}$) of the cavity using a vector network analyzer. Figures 2a and 2c display $S_{21}$ plotted against the probe frequency and the normalized flux bias $\phi = \Phi/\Phi_0$, where $\Phi$ represents the magnetic flux applied through the qubit loop and $\Phi_0$ represents the superconducting flux quantum.

For the conventional flux qubit, Fig. 2a shows clear anti-crossings located symmetrically on both sides of the $\phi = 0.5$ flux bias point, indicating that the qubit and cavity are coupled and their eigen-frequencies are matched. For the π-junction qubit, these anti-crossings appear symmetrically on both sides of the flux bias point $\phi = 0$, as well as $\phi = n$ with integer $n$, as shown in Fig. 2c, indicating half-flux-quantum shifted operation. At each optimal working flux bias point, marked by the dashed lines in Fig. 2, the cavity exhibits a fundamental resonance frequency of approximately $\omega_c/2\pi \sim 8.245$ GHz when coupled to the conventional flux qubit and 8.244 GHz when coupled to the π-junction qubit.

The transition frequency between the qubit ground and first excited states ($\omega_{01}/2\pi$) is shown in Figs. 2b and 2d. For the conventional flux qubit, at the flux-insensitive point (i.e., $\phi = 0.5$), the qubit has its minimum frequency at 4.173 GHz, detuned by $(\omega_c - \omega_{01})/2\pi = 4.072$ GHz from the fundamental cavity mode. The dash-dotted line in Fig. 2b is the theoretical fitting curve. Following ref. 13, the transition frequency of a C-shunt flux qubit is given by

$$\hbar\omega_{10} = \Delta + \frac{2\varepsilon^2}{\Delta} \quad (1)$$

where the energy gap $\Delta$ and the flux-dependent energy shift $\varepsilon$ are defined by

$$\Delta = \sqrt{4E_c E_J(1-2\alpha)} + \frac{8\alpha-1}{4(1-2\alpha)} E_c \tag{2}$$

$$\varepsilon = 2\sqrt{2}\pi\alpha E_J \left(\frac{E_c}{E_J(1-2\alpha)}\right)^{1/4} \left(\frac{\Phi}{\Phi_0} - 0.5\right) \tag{3}$$

Here, $E_c = e^2/2C_\Sigma$ is the effective charging energy determined by the total qubit capacitance $C_\Sigma$. Other parameters include the area ratio of the small JJ relative to the two larger JJs $\alpha$=0.458, Josephson energy $E_J/h$ = 225 GHz (where $h$ is Planck's constant, and the corresponding critical current density of the larger JJ in the qubit is 59 A cm$^{-2}$, which is close to the value obtained from junction characterization), and charging energy $E_C/h = (e^2/2C_\Sigma)/h$ = 0.130 GHz. The total qubit capacitance $C_\Sigma$ is 148 fF, which includes the shunt capacitance $C_S$ = 114 fF and the total junction capacitance of the flux qubit $C_J$ = 34.4 fF. Using the values of $E_c$ and $\alpha$, we estimate the qubit anharmonicity[13] $\hbar A = \frac{8\alpha-1}{4(1-2\alpha)} E_c$, defined as $(\omega_{12} - \omega_{01})/2\pi$ where $\omega_{12}/2\pi$ is the transition frequency between the qubit first and second excited states, to be 1.03 GHz. This value indicates a relatively strong anharmonicity in comparison with the 200-300 MHz typically observed in contemporary transmons. The detailed parameters can be found in the Methods section.

We note that the minimum in fig. 2d occurred at a finite current value of 56 μA. Considering that the period in the fig. 2c is 26.3 mA, we find that the offset flux is 0.002$\Phi_0$, which indicates the presence of a residual magnetic field. This residual field could be caused by components, such as circulators and isolators, that contain ferromagnetic materials. Another possible origin of the residual field is a spontaneous supercurrent at the 0-π phase boundary which can appear near the π-junction edges if the ferromagnetic layer thickness becomes smaller than the threshold needed for the π-state[28]. This situation can occur during the lift-off process and could be avoided by using dry etching.

For the π-junction qubit, at the flux-insensitive point (i.e., $\phi = 0$), the qubit has its minimum frequency at 4.055 GHz, detuned by $(\omega_c - \omega_{01})/2\pi = 4.189$ GHz from the fundamental cavity mode. The two qubits have similar transition frequencies because they have the same qubit design, except for the fact that one of them has a π-junction where the other one has only a shunting via. To explain why the conventional flux qubit and the π-junction qubit have similar frequencies despite their different circuit schemes, we examine the Lagrangian of the circuit in Supplementary Note 1. There, in the process of deriving the Hamiltonian of the π-junction qubit from the Lagrangian, for the π-junction with almost no phase change (i.e. the phase is fixed to π), the kinetic energy of the π-junction can be ignored. Thus, the Hamiltonian and the qubit frequency of the π-junction qubit is almost the same as those of the conventional flux qubits except for the phase shift of π in the potential energy. We believe that the small frequency difference between the two qubits can be attributed to the unavoidable small variation of $\alpha$ in the two qubits.

Despite the similar designs and qubit frequencies, the working flux bias points of the two qubits are entirely different. For the qubit with the π-junction, the optimal operation point, where the relevant transition frequencies are least sensitive to variations in the applied magnetic field, is at zero field, indicating a half-flux-quantum shifted operation compared to the conventional flux qubit. These results demonstrate the successful operation of the superconducting flux qubits operating at zero magnetic field when incorporating a π-junction.

**Energy relaxation time $T_1$ and dephasing time $T_2$.**

The coherence properties of the qubits were characterized by time-domain measurements, where the energy relaxation time ($T_1$) and spin-echo coherence time ($T_2$) were measured at the flux-insensitive point for each qubit using the control-pulse sequences depicted in the insets of Fig. 3. To measure $T_1$, we record the qubit's exited state population using a digitizer and plot the resulting signal as a function of the time delay (τ) in Figs. 3a and 3c. Figure 3a presents the energy relaxation data for the conventional flux qubit, which is well fitted by an exponential decay function, $\exp(-\tau/T_1)$, yielding $T_1 = 15.8 \pm 1.3\ \mu s$. For the π-junction qubit, a $T_1$ value of 1.45 ± 0.15 μs is observed, as shown in Fig. 3c.

The dephasing time, shown in Fig. 3b, d, was measured through spin-echo experiments. Using an exponential fit, we determine $T_2 = 11.3 \pm 0.76\ \mu s$ for the conventional flux qubit and $T_2 = 1.47 \pm 0.15\ \mu s$ μs for the π-junction qubit, at their respective optimal points (i.e., at $\phi = 0.5$ for the former and $\phi = 0$ for the latter).

Since $T_2$ is related to both the pure dephasing time $T_\varphi$ and $T_1$ as described by $\frac{1}{T_2} = \frac{1}{2T_1} + \frac{1}{T_\varphi}$, if the pure dephasing rate is negligible, $T_2$ becomes larger than $T_1$, approaching the $2T_1$ limit. Our current results for the coherence times of the conventional flux qubit ($T_1$ = 15.8 μs and $T_2$ = 11.3 μs) show $T_1 > T_2$, indicating the presence of substantial dephasing. The estimated $T_\varphi$ for the conventional flux qubit is about 17.6 μs. Usually, this tendency of $T_1 > T_2$ is common for conventional flux qubits (see Ref. 29). The shortened $T_2$ is often attributed to factors such as low-frequency charge two-level system (TLS) defects and spins.

In the π-junction flux qubit, $T_2 \approx T_1$. Also, both the $T_1$ and $T_2$ times of the π-junction qubit are one order of magnitude smaller than the respective times for the conventional flux qubit. From their coherence times ($T_1$ = 1.45 μs and $T_2$ = 1.47 μs), the $T_\varphi$ is estimated to be about 2.98 μs, which is shorter than that of the conventional flux qubit. These results indicate that additional dephasing source, such as inelastic quasiparticle tunneling through the metallic π-junction, causes $T_\varphi$ of the π-junction qubit to be shorter than that of the conventional flux qubit.

Furthermore, as shown in Figure 3c, d, the experimental data points exhibit large deviations from the fitting curve, indicating significant quasi-static fluctuations. Interestingly, these fluctuations appear to be larger at longer delays, which is unusual. This suggests that there might be an additional noise source affecting the data variability in π-junction qubit. One possible cause of this fluctuation could be magnetization fluctuations or slow movement of magnetic domains in the π-junction. However, further investigation is required to confirm this hypothesis.

We note that the coherence time of our conventional flux qubit was similar to that reported in Ref. 10 ($T_1 \sim 16\ \mu s$), where we used a transmission-line resonator to probe the qubit dispersively. We believe that our $T_1$ value is not limited by the measurement setup but by the dielectric dissipation arising from the material surface of the qubit or weakly coupled TLS defects in the remaining silicon dioxide after buffered hydrogen fluoride (BHF) treatment in our fabrication process.

**Quantitative analysis of coherence.**

It is worth noting that the coherence times of our π-junction qubit are three orders of magnitude higher than those of the previously demonstrated superconducting phase qubit coupled to a π-junction[19]. In reference 19, coherence times on the nanosecond scale indicate the presence of other dominant loss mechanisms. Therefore, it is difficult to determine the effect of the π-junction on the coherence time in that experiment. The microsecond-scale coherence times of our qubits made it possible for us to observe the effect of the π-junction and identify the presence of intrinsic decoherence in the π-junction.

On the other hand, the coherence times of the π-junction qubit were one order of magnitude lower than those of the conventional flux qubit. These results clearly show the impact of the NbN/PdNi/NbN π-junction on both $T_1$ and $T_2$, providing valuable insight into the coherence properties of the superconducting flux qubits. In particular, we expect that the qubit coherence is limited by the π-junction.

To understand the main factors contributing to the reduced coherence times of the π-junction qubit, we estimated the theoretical predictions for the decay time caused by damping at the π-junction as discussed in Refs. 19, 30. Since the superconductor/ferromagnetic metal/superconductor π-junction can be considered a superconductor/normal metal/superconductor (S/N/S) structure, and the latter is known to have dissipation via gapless quasiparticle excitations, a similar theoretical model of dissipation effects in the ferromagnetic metal layer can be utilized. In order to describe the damped dynamics for the π-junction qubit, the resistively shunted junction (RSJ) model for the π-junction, where dissipation occurs in the normal resistance $R_{N,\pi}$ of the junction, was used in Ref. 30. Our NbN/PdNi/NbN π-junction is an overdamped junction and has $R_{N,\pi} \approx 344.3$ μΩ and a critical current $I_{c,\pi} \approx 3.5$ mA which are estimated from I-V characteristics measured at 4.2 K (see Supplementary Figure 1 and Supplementary Note 2). The obtained current density of the π-junction was 4.4 kA cm$^{-2}$. In our case, the condition of the qubit level splitting $\Delta \gg 2eI_{c,\pi}R_{N,\pi}$ is satisfied, allowing us to use a simple approximate expression for the relaxation time $\tau_{relax} \approx \frac{\Delta}{2I_c^2 R_{N,\pi}}$,[19, 30] where $\Delta \approx h \cdot (4\ \text{GHz})$ (with e being the elementary charge). Here the energy $2eI_{c,\pi}R_{N,\pi} \approx h \cdot (583\ \text{MHz})$ is associated with the characteristic Josephson frequency of our NbN/PdNi/NbN π-junction. Based on these calculations, we theoretically estimate the relaxation time to be approximately $\tau_{relax} \approx 87.3$ ns. Here $I_\text{c} \approx 210$ nA is the critical current of the small NbN/AlN/NbN qubit JJ (taking into account the current density obtained from the fitting parameter of qubit spectrum (59 A cm$^{-2}$) and the reduced JJ diameter due to the etching process to be 0.67 µm). It is worth noting that this estimated value is significantly lower than the experimentally measured value. Nevertheless, it clearly demonstrates the additional decoherence induced by the π-junction employed in this flux qubit, consistent with our findings. We therefore attribute the difference between the qubits with and without π-junctions to dissipation caused by quasiparticles in the superconductor/ferromagnetic metal/superconductor structure in the π-junction qubit. Even if we want to keep using an overdamped π-junction, it could be possible to increase $\tau_{relax}$ by reducing the JJ size, which reduces $I_\text{c}$, and/or increasing the π-junction size, which reduces $R_{N,\pi}$. However, Ref. 30 suggested that using an underdamped π-junction (expected from the junction with insulating tunnel barrier) is a more promising approach to give a significantly longer coherence time. In other words, the coherence properties of our π-junction qubit could be improved by employing a ferromagnetic insulator in the π-junction.

## Conclusions

We have realized a superconducting flux qubit operating at zero magnetic field by utilizing a ferromagnetic Josephson π-junction. We engineered the NbN/PdNi/NbN π-junction incorporated into NbN/AlN/NbN-based superconducting flux qubits with a nanoscale thickness in the range needed to produce a robust π-state. The microwave spectroscopy and time-domain coherence measurements of the π-junction qubit confirmed the optimal operation at zero magnetic field.

The qubit lifetime is in the microsecond range. The lifetime of the π-junction qubit is an order of magnitude shorter than that of a reference conventional flux qubit. We attribute the difference to dissipation caused by quasiparticles in the superconductor/ferromagnetic metal/superconductor (S/FM/S) structure in the π-junction qubit, indicating that the coherence properties could potentially be improved by employing a ferromagnetic insulator in the π-junction.

In this work, we prioritized implementing a flux qubit with a well-established S/FM/S structure that reliably achieves a stable π-state[24] and focused on demonstrating flux bias-free operation. This study is highly significant, as it experimentally demonstrates a flux qubit with an SFS π-junction operating without flux bias. Based on the π-junction qubit fabrication techniques established in this experiment, we are considering incorporating recent advances such as S/I/F/S[31-34] and S/ferromagnetic insulator (FI) (e.g., gadolinium nitride GdN)/S junctions[35, 36] in future research. This approach holds promise for further improving the performance and deepening our understanding of qubit systems.

Unlike the phase qubit studied in Ref. 19, reporting clear coherence properties such as $T_1$ and $T_2$ for a flux qubit that incorporates a π-junction represents a significant advance. Additionally, while Shcherbakova and colleagues have attempted to implement π-junctions in flux qubits (in Ref. 20), they have not demonstrated flux bias-free operation or measured $T_1$. Therefore, experimentally demonstrating flux bias-free operation of a flux qubit and the clear role of the π-junction in determining the coherence properties is a significant step. By fabricating the π-junction using the same NbN-based electrodes in the qubit structure, we are able to construct monolithic quantum circuits, which may have advantages for large-scale integrated circuits.

This qubit can open the pathway towards high density integration of flux qubits. The incorporation of π-junction flux qubits in the 3D architecture can also be utilized for bosonic code quantum computation[37]. It is also worth noting that the use of ferromagnetic junctions in qubits can be useful for alternative control (and therefore alternative layouts) of qubit operations[38-40]. With further material improvements this qubit can also be a long coherence qubit for quantum computing and a highly sensitive nanoscale magnetic field sensor.

## Methods

**Fabrication of all-nitride flux qubit with NbN/PdNi/NbN π-junction.**

In this study, we utilized two distinct types of JJs as the building blocks for superconducting flux qubits. The first type is a fully epitaxial NbN/AlN/NbN JJ grown on a single-crystal Si (100) substrate, with a (200)-oriented TiN buffer layer (see the left inset of Fig. 1c). The second type is a NbN/PdNi/NbN ferromagnetic JJ designed as a π-phase shifter, fabricated on the NbN/TiN layers on the same substrate (see the left inset of Fig. 1d). While NbN-based π-junctions with CuNi interlayers are well established as π-phase shifters[41,42], we adopted the PdNi interlayer due to its significantly smaller spin-flip scattering, resulting from a lower density of Ni magnetic clusters compared to the CuNi interlayer. This improves magnetic uniformity and the longer decay length[24], making PdNi a suitable choice for device applications by providing better control of critical currents and ensuring functional π-junctions. Further details about the fabrication process for each junction structure can be found in our earlier reports[24, 43, 44].

To examine the effect of the ferromagnetic JJ as a π-phase shifter on the qubit, we fabricated flux qubits with and without π-junctions on the same Si substrate, as shown in Fig 1. The dimensions of the qubit chip are 2 mm × 7 mm × 0.45 mm. In summary, the fabrication process of our samples involved the following steps as shown in Fig. 4: (a) After surface cleaning of a 2-inch Si wafer, (b) a 50 nm-thick TiN layer was grown on it using DC reactive sputtering. (c) Next, a tri-layer structure consisting of NbN (100 nm) / AlN (~1.8 nm) / NbN (200 nm) was deposited using DC reactive sputtering. (d) The first patterns, the JJs for the qubit, were defined using electron-beam lithography and reactive ion etching (RIE) using tetrafluoromethane ($CF_4$) gas for NbN and argon (Ar) gas for AlN. In this step, the top two layers (NbN/AlN) were removed except in the JJ region. The bottom NbN layer is left almost intact. (e) Subsequently, the shunt capacitor, shown as two rectangular pads in Fig. 1(b), and the qubit base electrodes, were formed through photolithography using an i-line stepper and RIE. (f) For the π-junction, a ferromagnetic JJ was additionally fabricated using photolithography and a lift-off process. This process entailed creating a circular junction of NbN (93 nm) / PdNi (15 nm) / NbN (93 nm) with a diameter of 10 μm, as shown in Fig. 1(d), on one of the base electrodes of the qubit. Importantly, the PdNi interlayer thickness was set to 15 nm to achieve both the π-state and a critical current density ($J_c$) ~ 4.4 kA cm$^{-2}$, sufficient to function as a π-phase shifter for the qubit[24]. Considering that $Pd_{89}Ni_{11}$ in the thickness range 8-20 nm corresponds to the π-state[24], the 15 nm thickness of the PdNi layer in our π-junction is near optimal to obtain a robust π-state, even in the presence of fluctuations in film thickness uniformity especially at the pattern edge that forms during the lift-off process. Additionally, it is worth mentioning that the total thickness of the ferromagnetic JJ was intentionally set to be comparable to the total height of the JJ for subsequent planarization. (g) After depositing a silicon dioxide ($SiO_2$) film as an insulating layer between the base and wiring layers, we performed chemical mechanical polishing (CMP) for planarization. (h) Following the planarization process, the contact via-holes were patterned by an i-line stepper followed by an RIE using trifluoromethane ($CHF_3$) gas. (i) The upper wiring layer was prepared using a 300 nm-thick NbTiN film deposited by DC magnetron sputtering, followed by photolithography and RIE using $CF_4$ gas. (j) Finally, to avoid unwanted TLS in $SiO_2$, all $SiO_2$ film were removed by BHF etching.

**Junction parameters.**

By measuring the current-voltage characteristics of test JJs fabricated on the same wafer as the qubits at 4.2 K, we found that the $J_c$ was approximately 60 - 65 A $cm^{-2}$ for the NbN/AlN/NbN JJs and about 4.4 kA $cm^{-2}$ for the NbN/CuNi/NbN ferromagnetic JJs. The $J_c$ of the π-junctions is much higher than that of the conventional JJs in the qubit so that the π-junction remains in a superconducting state and is operated in the regime of a well-defined phase. We also confirmed that similar π-junctions, with the same 15 nm-thick PdNi layer, effectively induced a half-flux-quantum shift in SQUID structures on a test chip before qubit measurement (cf. Ref. 24). The magnetic field-dependence of the measured SQUID critical current showed that our NbN/PdNi/NbN junctions were in the π-state, as described in Ref. 24.

Figure 1c illustrates the three-junction type of the flux qubit[45], which consists of a superconducting loop with three Josephson junctions. Here, the third junction in the flux qubit loop has a smaller area and controls the anharmonicity. The ratio between the size of the smaller junction and the bigger ones, i.e. the parameter *α*, is typically chosen such that the effective potential energy term for the flux variable is a double-well potential, as opposed to the single-well potential that is typical for phase qubits. The two larger JJs were designed to have 1.0 μm diameter (using a mask size of 1.2 μm diameter and expecting a reduction of 0.20 μm after the fabrication process), and the third JJ was designed to have a 0.68 μm diameter (using a mask size of 0.88 μm) to get a smaller area by a factor $\alpha \approx 0.46$. The best fitting parameters for the qubit spectrum give 0.99 μm for the diameter of the larger JJs, 0.67μm for the diameter of the smaller JJ, i.e., $\alpha \approx 0.46$, and a reduction of 0.21 μm compared to the JJ diameters in the qubit design. The capacitance of the larger JJ in the flux qubit is $C_J$ = 36.0 fF and that of the smaller JJ is $\alpha C$ which is estimated in the same manner as in Ref. 9 and the qubit's shunt capacitance is calculated to be about 123 fF by sonnet simulation.

**Experimental setup.**

Experiments were performed in a dilution refrigerator with a base temperature of 10 mK. The qubit chip was mounted in a 3D microwave cavity, as shown in Fig. 1a. The cavity was attached to a cold finger of the dilution refrigerator and covered by a three-layer shield consisting of one aluminum-based superconducting shield and two μ-metal magnetic shields[13]. Inside the dilution refrigerator, microwave lines were carefully filtered, attenuated and isolated.

To characterize the cavity, the microwave transmission $S_{21}$ was measured using a vector network analyzer. For spectroscopy and coherence measurements of the qubit, an additional microwave drive and a commercial analogue-to-digital converter were used, enabling the qubit state to be read out dispersively via the cavity in a circuit quantum electrodynamics (circuit QED) architecture. To apply the magnetic flux bias to the qubit, a small custom-made solenoid magnet outside the 3D cavity was utilized. A detailed description of the experimental setup can be found in Ref. 13.

## Data availability

The data that support the findings of this study are available from the corresponding authors upon reasonable request.

## Acknowledgments (not compulsory)

This work was supported by Japan Science and Technology Agency Core Research for Evolutionary Science and Technology (Grant No. JPMJCR1775), JSPS KAKENHI (JP19H05615), JST ERATO (JPMJER1601), and partially by MEXT Quantum Leap Flagship Programs (JPMXS0120319794 and JPMXS0118068682). We are grateful to Alexey K. Feofanov and Akira Fujimaki for fruitful discussions about π-junctions. We also acknowledge Imran Mahoob for his helpful contribution regarding the measurement setup. S.K. acknowledges Kazuyo Takaki for her help with sample fabrication. T.Y. acknowledges Center for Heterogeneous Quantum/Material Fusion Technologies, Center for Key Interdisciplinary Research, Tohoku University.

## Author contributions statement

All authors contributed extensively to the work presented in this article. S.K., T.Y., H.T., and K.S. designed the experiment. S.K. designed the samples. D.P., W.Q., H.T. and T.Y. fabricated the samples and characterized the basic junction properties. L.V.A. and S.S. provided the measurement setup and performed the microwave measurements. S.K. analyzed the data and wrote the manuscript with feedback from all authors. L.V.A. and S.A. contributed to data analysis. K.S. supervised the project.

## Additional Information
### Competing interests
The authors declare no competing interests.

### Corresponding authors
Correspondence and requests for materials should be addressed to S.K, L.V.A. or T.Y.

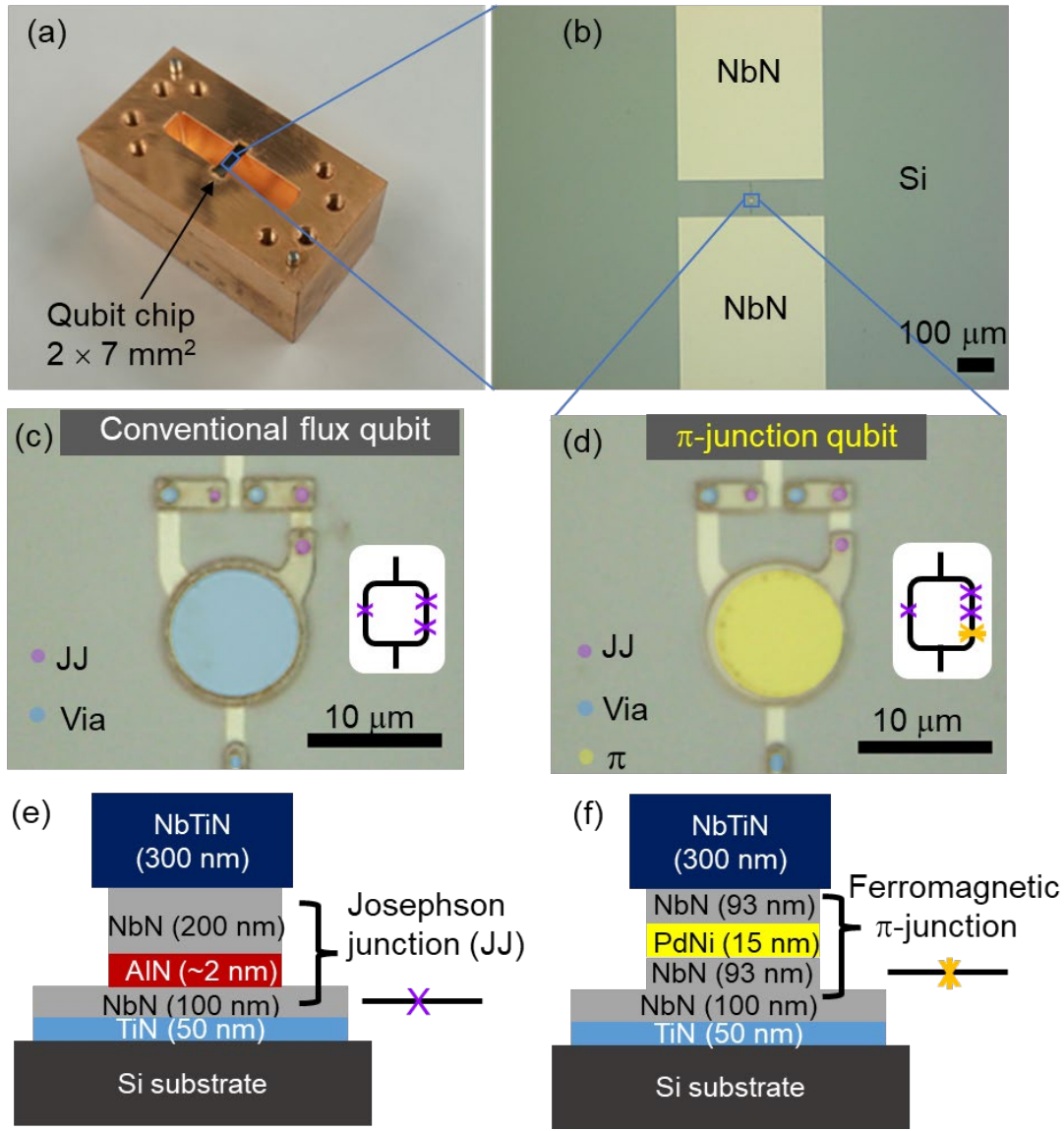

**Figure 1. NbN-based flux qubits and 3D cavity.**
(**a**) Photograph of the 3D cavity used for qubit measurement. The mounted chip size is 5 × 7 mm². (**b**) Optical microscope image of capacitively-shunted flux qubit made of nitride-based superconductors. Two rectangular capacitors with sizes of 400 × 500 μm² are coupled to shunt the central qubit circuit. (**c**) False-color image of a conventional flux qubit used as a reference device. (**d**) False-color image of a flux qubit that incorporates a π-junction. Here, the Josephson junction (JJ), Via-hole (Via), and π-junction (π) are marked by purple circles, blue circles, and yellow circle, respectively. The insets in (**c**, **d**) display circuit diagrams of the flux qubit loop with JJs (marked by ×) and π-JJ (∗). (**e**, **f**) Schematic views of junction cross-section structures for (**e**) the Josephson junctions and (**f**) the ferromagnetic π-junction.

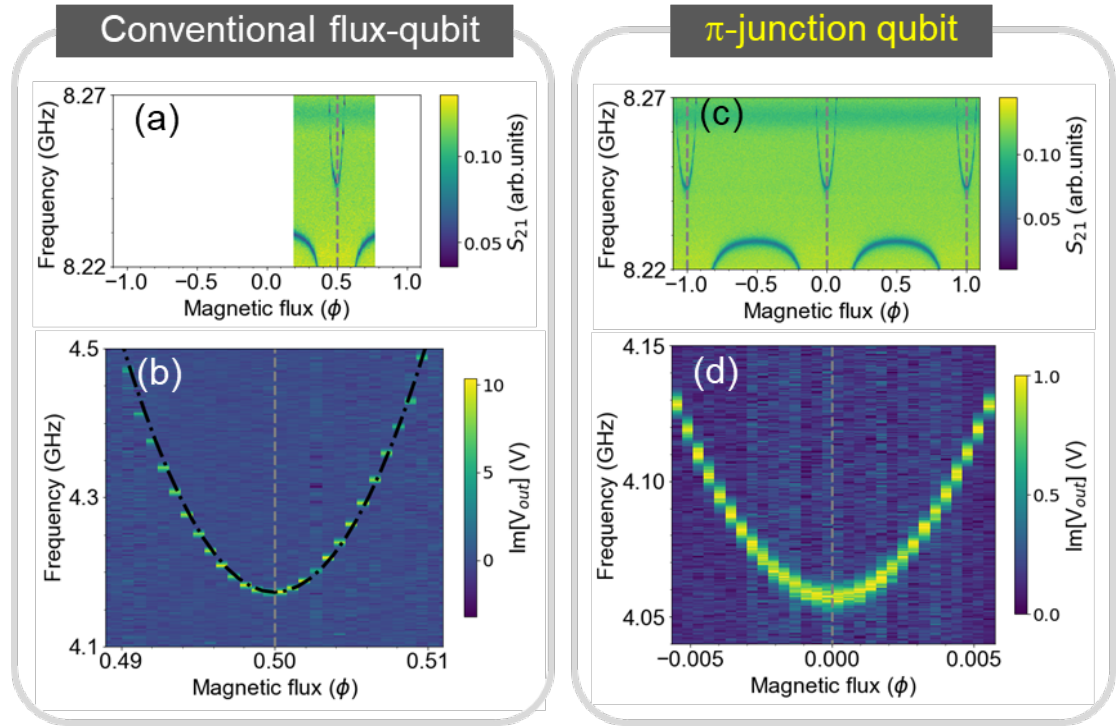

**Figure 2. Spectroscopy of cavity and qubit**.
(**a**) Cavity spectrum corresponding to microwave transmission ($S_{21}$) of the 3D cavity coupled to the conventional flux qubit as a function of probe frequency and normalized magnetic flux ($\phi$). (**b**) Conventional qubit spectrum for the transition from the ground state to the first excited state using dispersive readout. The black dash-dotted line corresponds to the theoretical fitting curve. (**c**) Cavity spectrum and (**d**) qubit spectrum for the case of the π-junction qubit. The dashed lines mark the flux values for which the qubits have their minimum transition frequencies.

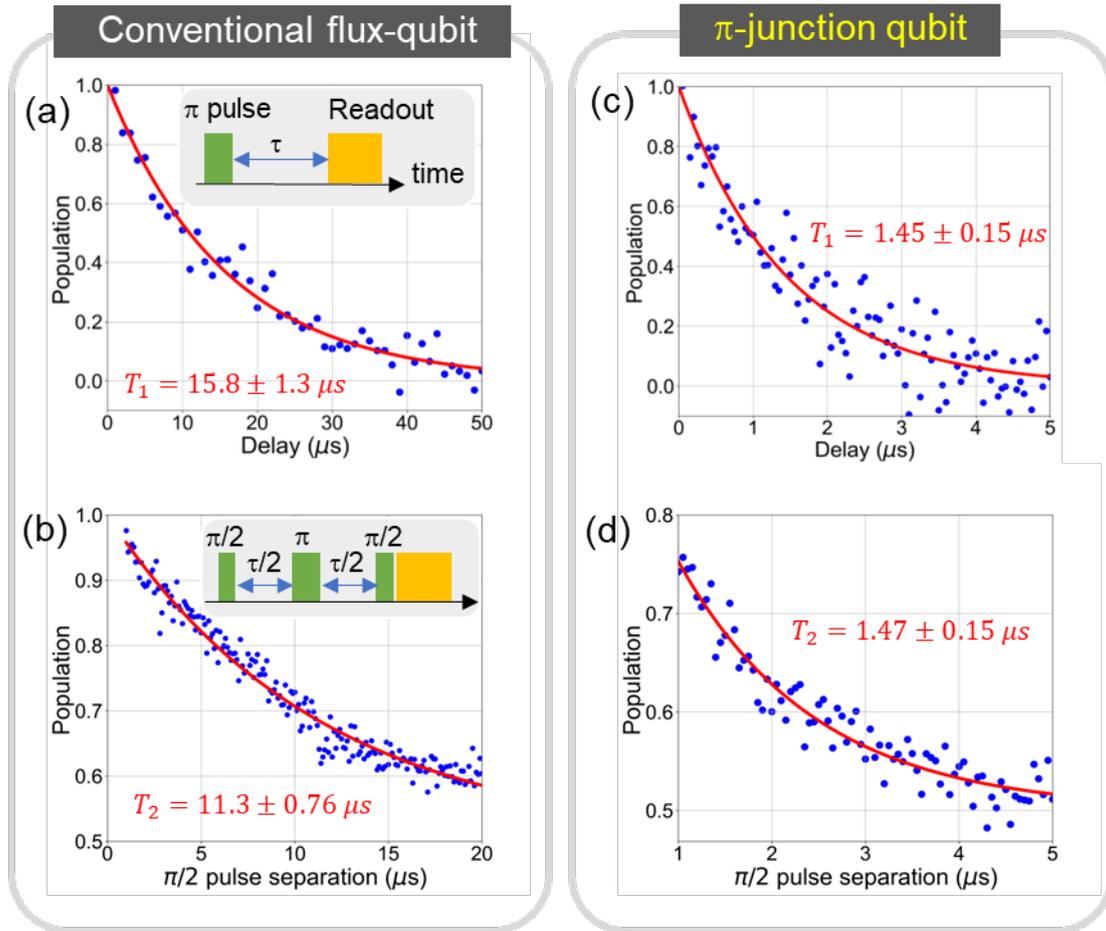

**Figure 3. Qubit coherence characterization.**
Decay profiles of (**a**) energy and (**b**) phase coherence signals for conventional flux qubit. The exponential fits (solid red lines) give coherence times of $T_1 = 15.80 \pm 1.30$ μs and $T_2 = 11.34 \pm 0.76$ μs. The insets show schematic diagrams of the corresponding measurement pulse sequences. (**c**) and (**d**) show similar signals for the $\pi$-junction qubit, with coherence times of $T_1 = 1.45 \pm 0.15$ μs and $T_2 = 1.47 \pm 0.15$ μs.

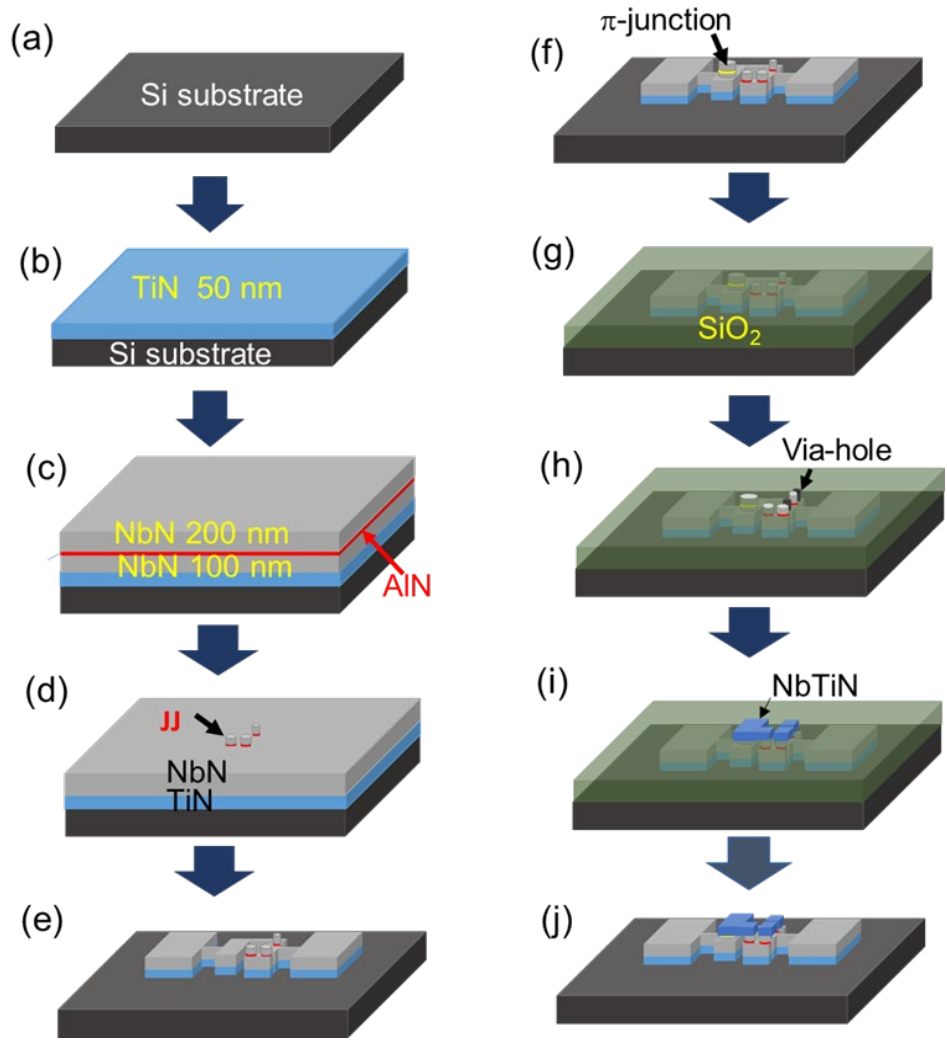

**Figure 4. Fabrication process flow schematics of NbN-based superconducting flux qubit with a ferromagnetic Josephson π-junction.**
(**a**) Surface cleaning of the Si substrate, (**b**) Epitaxial film growth of a TiN film as a buffer layer on the Si substrate. (**c**) Epitaxial growth of NbN/AlN/NbN tri-layer. (**d**) Patterning of the JJ using electron-beam lithography and RIE. (**e**) Patterning of base electrodes including the shunt capacitor. (**f**) Formation of the π-junction by deposition of the NbN/PdNi/NbN tri-layer, followed by patterning using photolithography and a lift-off process. (**g**) Deposition of $SiO_2$ and planarization by chemical mechanical polishing. (**h**) Formation of via-hole using photolithography and RIE. (**i**) Deposition and patterning of NbTiN for the upper wiring layer. (**j**) Removal of $SiO_2$ using BHF treatment.

# Supplementary Information:
# Superconducting flux qubit with ferromagnetic Josephson π-junction operating at zero magnetic field

Sunmi Kim[1, *, †], Leonid V. Abdurakhimov[2, 5, ‡, †], Duong Pham[3, 6], Wei Qiu[4, 5], Hirotaka Terai[4], Sahel Ashhab[1], Shiro Saito[2], Taro Yamashita[3, 6, §], and Kouichi Semba[1,7]

[1]Advanced ICT Research Institute, National Institute of Information and Communications Technology (NICT), 4-2-1, Nukui-Kitamachi, Koganei, Tokyo 184-8795, Japan

[2]NTT Basic Research Laboratories, NTT Corporation, 3-1 Morinosato-Wakamiya, Atsugi, Kanagawa 243-0198, Japan

[3]Graduate School of Engineering, Nagoya University, Furo-cho, Chikusa-ku, Nagoya, Aichi 464-8603, Japan

[4]Advanced ICT Research Institute, NICT, 588-2 Iwaoka, Nishi-ku, Kobe, Hyogo 651-2492, Japan

[5]present address: IQM Finland Oy, Keilaranta 19 02150, Espoo, Finland

[6]present address: Graduate School of Engineering, Tohoku University, Aoba 6-6-05, Aoba-ku, Sendai, 980-8579, Japan

[7]present address: Institute for Photon Science and Technology, The University of Tokyo, Tokyo 113-0033, Japan

[*]kimsunmi@nict.go.jp, [‡]leonid.abdurakhimov@meetiqm.com, [§]taro.yamashita.c8@tohoku.ac.jp

[†]These authors contributed equally to this work.

## Supplementary Note 1. Lagrangian of π-junction qubit

Let us take the qubit loop shown in the inset of Fig. 1(d). Since the Lagrangian of each junction is defined as the difference between the kinetic energy ($T = \frac{C\phi_0^2}{2}\dot{\delta}^2$) and potential energy terms ($U = E_J(1 - \cos\delta)$), the Lagrangian of the whole circuit can be written as

$$\mathcal{L} = T - U = \sum_{i=1}^{3}\left(\frac{C_i\phi_0^2}{2}\dot{\delta}_i^2 - E_{J,i}(1 - \cos\delta_i)\right) + \frac{C_\pi\phi_0^2}{2}\dot{\delta}_\pi^2 - E_{J,\pi}(1 - \cos(\delta_\pi + \pi)) \tag{S1}$$

where $C_i$ is the capacitance, $\phi_0 = \Phi_0/2\pi$, $\Phi_0$ is the superconducting flux quantum, $E_{J,i}$ is the Josephson energy given by $E_{J,i} = I_{c,i}\Phi_0/(2\pi)$, $I_{c,i}$ is the critical current, and $\delta_i$ is the dimensionless phase variable. Here, the index $i$ labels the three conventional Josephson junctions ($i$ = 1, 2, 3), as well as π for the π-junction. Importantly, the terms that correspond to the π-junction look almost identical to those of the other junctions, except for the π-phase shift in the potential energy term. Flux quantization leads to the constraint:

$$\sum_{i=1}^{3}\delta_i + \delta_\pi = \frac{2\pi\Phi_{ext}}{\Phi_0} + 2\pi n \tag{S2}$$

where $n$ is an integer. In principle, to analyze the quantum dynamics of the different variables, we need to substitute the constraint into the Lagrangian, calculate the Hamiltonian via a Legendre transformation, introduce the quantum commutation relations and then analyze the properties and dynamics of different variables. However, we can already say something about the behavior of the variables based on the Lagrangian in Eq. (S1).

The classical ground state of the system is obtained by minimizing the potential energy. Since the π-junction's critical current is four orders of magnitude larger than those of the other junctions, its potential energy term will dominate in comparison to the other junctions. The classical ground state must therefore have $\delta_\pi \approx \pi$. Furthermore, since the capacitance $C_\pi$ is comparable to those of the conventional Josephson junctions[1], the strong confinement of the variable $\delta_\pi$ implies that, to a very good approximation, quantum fluctuations in $\delta_\pi$ are negligible compared to the other phase variables $\delta_i$. Regarding the $C_\pi$ of our SFS junction, we assume that it is of the same order of magnitude as that in Ref.1. In the Supplementary Information of Ref. 1, Feofanov *et al.* estimated the capacitance $C_\pi$ of their SFS junction to be less than 10 fF. Considering the capacitance of the larger Josephson junction (JJ) ($C_J$ = 36.0 fF) and the smaller JJ ($\alpha C \approx$ 16.5 fF) in our flux qubit, we estimate that $C_\pi$ of our SFS junction is comparable to those of the other junctions.

As a result, we can set $\delta_\pi = \pi$ and $\dot{\delta}_\pi = 0$ to obtain the approximate Lagrangian

$$\mathcal{L} = \sum_{i=1}^{3}\left(\frac{C_i\Phi_0^2}{2}\dot{\delta}_i^2 - E_{J,i}(1-\cos\delta_i)\right) \tag{S3}$$

with the flux quantization condition

$$\sum_{i=1}^{3}\delta_i = \frac{2\pi\Phi_{ext}}{\Phi_0} + 2\pi n - \pi \tag{S5}$$

In other words, the Lagrangian of the π-junction flux qubit reduces to that of an equivalent circuit without the π-junction, but with half a flux quantum added to the applied flux bias.

## Supplementary Note 2. Estimation of $I_c$ and $R_n$ of π-junction

For the NbN/PdNi/NbN junction with a 10 μm-square-sized junction, we measured I-V characteristics at 4.2 K as shown in the following figure. Their critical current ($I_c$) and the normal resistance ($R_n$) are about 4.4 mA and 270.3 μΩ. The current density ($J_c$) of the π-junction is estimated to be about 4.4 kA cm$^{-2}$. Considering a circular shape with a diameter of 10 μm used for a π-junction in qubit, the extracted $I_c$ and $R_n$ are 3.5 mA and 344.3 μΩ.

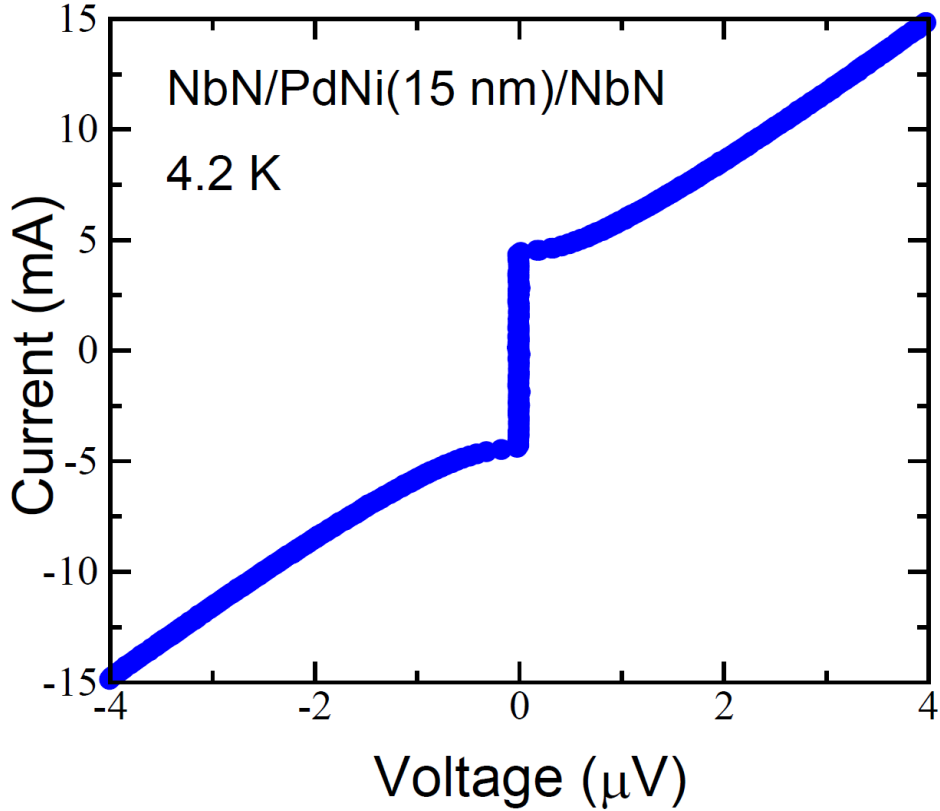

**Supplementary Figure 1.** Current-volage (I-V) characteristics of a 10 × 10 μm$^2$ π-

junction with a 15 nm-thick PdNi interlayer measured at 4.2 K.

## Supplementary References